\documentclass[12pt]{article}
\usepackage{amsmath}
\begin{document}

\title{The Case for 2-D Turbulence in Antarctic Data}

\author{Mayer Humi\\
Department of Mathematical Sciences\\
Worcester Polytechnic Institute\\
100 Institute Road\\
Worcester, MA  0l609}

\maketitle
\thispagestyle{empty}

\newpage

\begin{abstract}

In this paper we examine the data that was collected at Haley Station
in Antarctica on June 22, 1987.  Using a test devised by Dewan \cite
{EMD} we interpret the flow as one which represents two-dimensional
turbulence.  We also construct a model to interpret the
spectrum of this data which is almost independent of the wave number
for a range of frequencies.
\end{abstract}

\newpage

\section{Introduction}

Two dimensional turbulence has been the subject of intense theoretical
research \cite{VMC,MEM} and simulation experiments \cite{WH}.  The reason
for this interest stems from the fundamental differences between 3-d
isotropic and 2-d turbulence.  To begin with, vortex stretching is
absent in 2-d as a direct consequence of Navier-Stokes equations.
Furthermore in 3-d the energy cascade is from the large eddies to small
one but this process reverses itself in 2-d and leads to the formation
of large scale coherent eddies.  Another difference between two and three dimensional
turbulence exists in the inertial range of the spectrum.  Kraichnan
showed \cite{RK} that in 2-d in addition to Kolmogorov inertial range
there is (due to ensotrophy conservation in zero viscosity) another
scaling law in the form
$$
E(k) = c\eta^{2/3}k^{-3}
$$
where  $\eta$  is ensotrophy dissipation rate.

While many simulations \cite{DKL,GKB} confirm these theoretical predictions
the actual observation and detection of 2-d turbulence as a natural
phenomena remains (as far as we know) an open questions.

One of the objectives of this paper is to weigh in the pros and cons
for 2-d turbulence in the Antarctic data that was obtained by the
British observation post as Haley Station in Antarctica on June 22, 1987 (for
further description of this data see \cite{JCK,NRE}).  The importance of
these measurements stem from the fact that the flow field  ${\bf u} =
(u,v,w)$  and the temperatures were measured simultaneously at three
different heights viz. 5m, 16m and 32m.  These simultaneous readings
enable us to apply a test devised by E. Dewan \cite{EMD} for the
detection of 2-d turbulence.  According to this test 2-d turbulence is
characterized by small values for the coherence \cite{WNV} between
the time series which represent the various meteorological variables
at different heights.

From another point of view the Antarctic data represent a stably
stratified medium.  (According to mission records the temperature
gradient with height can reach up to $1 K/m$).  Under these
circumstances Bolgiano \cite{RBJ,BRJ} and others \cite{EMD} speculated
about the existence of ``buoyancy range turbulence'' (BRT) which
should lead to a flattening of the spectra in parts of the inertial
range.  In this paper we shall estimate the power spectrum for the
data using the usual Fourier transform and by the method of maximum
entropy (briefly the reason for this duplicatcy is due to the
existence of ``discontinuities'' in the data).  Both of these
estimates show a spectral range in which the spectrum is almost flat
and thus support the theoretical arguments that were advanced for the
existence of BRT.

The plan of the paper is as follows:  In section 2 we describe the
method used to filter out the mean flow and waves from the data and
the tests that were applied to verify that the residuals actually
represent turbulence.  In section 3 we apply the coherence test for
2-d turbulence and discuss its consequences.  In section 4 we present
a model for the power spectrum of the data and its implications.  We end up in
section 5 with some conclusions.

\section{Data Detrending}

The statistical approach to turbulence splits the flow variables
$\tilde{\bf u}, \tilde{T}$  (where $ \tilde{T}$  is the temperature)
into a sum
$$
\tilde{\bf u}= {\bf u} + {\bf u}^\prime + {\bf u},\;\;\tilde{T} =
T + T^\prime + t
$$
where  ${\bf u}, T$  represent the mean (large scale) flow, $u^\prime,
T^\prime$  represent waves and  $u,t$ ``turbulent residuals''
\cite{JJF}

To effect such a decomposition in our data we used the Karahunan-Loeve (K-L)
decomposition algorithm (or PCA) which was used by many researchers
(for a review see \cite{CPMG}).  Here we shall give only a brief
overview of this algorithm within our context.

Let be given a time series  $X$  (of length  $N$) of some geophysical
variable.  We first determine a time delay  $\Delta$  for which the
points in the series are decorrelated.  Using  $\Delta$  we create
$n$  copies of the original series
$$
X(k),\;\;X(d + \Delta),\ldots,X(k + (n - 1)\Delta).
$$
(To create these one uses either periodicity or choose to consider
shorter time-series).  Then one computes the auto-covariance matrix
$R = (R_{ij})$
\begin{equation}
\label{Delta}
R_{ij} = \displaystyle\sum^N_{k=1} X(k + i\Delta)X(k+j\Delta).
\end{equation}
Let  $\lambda_0 > \lambda_1, \ldots, > \lambda_{n-1}$  be the
eigenvalues of  $R$  with their corresponding eigenvectors
$$
\phi^i = (\phi^i_0, \ldots, \phi^i_{n-1}),\;\;i = 0, \ldots, n - 1.
$$
The original time series  $T$  can be reconstructed then as
\begin{equation}
\label{phi}
X(j) = \displaystyle\sum^{n-1}_{k=0} a_k(j)\phi^k_0
\end{equation}
where
\begin{equation}
\label{sum}
a_k(j) = \displaystyle\frac{1}{n}\displaystyle\sum^{n-1}_{i=0}X(j +
i\Delta)\phi^k_i.
\end{equation}
The essence of the K-L decomposition is based on the recognition that
if a large spectral gap exists after the first  $m_1$  eigenvalues of
$R$  then one can reconstruct the mean flow (or the large component
( of the data by using only the first  $m_1$  eigenfunctions in
(\ref{phi}).  A recent refinement of this procedure due to Ghil et al
\cite{CPMG} is that the data corresponding to eigenvalues between  $m_1
+ 1$  and up to the point  $m_2$  where they start to form a
``continuum'' represent waves.  The location of  $m_2$ can be
ascertained further by applying the tests devised by Axford \cite{DNA}
and Dewan \cite{EMD} (see below).

Thus the original data can be decomposed into mean flow, waves and
residuals (i.e. data corresponding to eigenvalues  $m_2 + 1,\ldots, n
- 1$  which we wish to interpret at least partly as turbulent
residuals).

For the data under consideration we carried out this decomposition
using a delay  $\Delta$ of 1024 points (approximately 51 sec.)
for all the geophysical variables.  In table 1 we present the values
of  $m_1, m_2$  that were used in this decomposition for the flow
variables at different heights.  (In all cases  $n = 64$).

The residuals of the time series which are reconstructed as
\begin{equation}
\label{ak}
X^r(j) = \displaystyle\sum^{n-1}_{k=m_2+1} a_k(j)\phi^k_0
\end{equation}
contain (obviously) the measurement errors in the data.  However to
ascertain that they should be interpreted primarily as representing
turbulence we utilize the tests devised by Axford \cite{DNA} and Dewan
\cite{EMD}.  According to these tests turbulence data (at the same
location) is characterized
by low coherence between  $u,v,w$ and a phase close to zero or  $\pi$
between  $w$  and  $t$.  (A phase close to  $\pi/2$  is
characteristic of waves).  Figs. 1,2,3 show samples of the coherence
between the residuals of  $u,v,w$ at different heights.  They
demonstrate that for most frequencies the coherence is less than 0.1.  Fig. 4 gives a
scatter plot of the phase between $w$ and  $t$ at height 5m.  This figure is less
definitive as there are still quite a few points in the wave sector
$\left(\displaystyle\frac{\pi}{4},\;\displaystyle\frac{3\pi}{4} \right)$.  However out of
the 200 points in this plot 125 are in the ``turbulence sector''.

These tests show that to a large extent the residuals that were
obtained from the K-L decomposition represent actual turbulence.

\section{Tests for 2-d turbulence}

In today literature \cite{UF} a spectral slope of $-3$ in part of the
inertial range is considered to be a strong indicator for 2-d turbulence.  However as noted already by Lily \cite{DKL} ``geophysical
consideration'' might modify this slope.  Since the spectral plots for
the flow under consideration (for sample see figs. 8,9,10) do not exhibit this
dependency (except for $w$ at 16m in the low frequencies) we must resort to other tests to bolster the claim that the
flow described by this data corresponds to 2-d turbulence.

To this end we utilize a test devised by Dewan \cite{EMD}.  According
to this test inviscid two dimensional turbulence is characterized by
the fact that the temporal statistical coherency \cite{WNV} between
the time series representing the flow variables at different altitudes
is zero.  With viscosity taken into account some vertical separation of the order of (10m for air) is needed for the coherency to become
small. (Strong coherency with values close to one indicates a strong
linear relationship between the two time series \cite{WNV}). 

Some typical plots for the coherency in the data is presented in
figs. (5,6,7).  In these plots the coherency for $w$ between the
different heights is plotted for different wave numbers.  We observe that
for most sampled frequencies the coherency is well below 0.1 and according to Dewan \cite{EMD}
``these values constitute evidence for 2-d turbulence and against
other types of fluctuations''.

\setcounter{equation}{0}

\section{The spectrum}

Two dimensional flow of incompressible and inviscid fluid conserve
both the energy  $E$  and the enstrophy  $\Omega$.  Fir viscous fluid
these quantities decay according to
\begin{equation}
\label{epsilon}
-\epsilon = \displaystyle\frac{\partial E}{\partial t} = -2 \nu
 \Omega,\;\;-\epsilon_\omega =
 \displaystyle\frac{\partial\Omega}{\partial t} =
 -\nu\mid\overline{\nabla \omega}\mid^2
\end{equation}
The energy spectrum is determined therefore by both parameters
$\epsilon, \epsilon_\omega$ which leads to the definition of a length
scale
\begin{equation}
\label{omega}
L_\omega =
\left(\displaystyle\frac{\epsilon}{\epsilon_\omega}\right)^{1/2}
\end{equation}
From dimensional considerations one concludes then that \cite{ASM} the
energy spectrum in the inertial range must have the form
\begin{equation}
\label{range}
E(k) = f(kL_\omega)\epsilon^{2/3} k^{-5/3}
\end{equation}
where  $f$  is a function of the dimensionless variable  $kL_\omega$.
If at one end of the inertial range only  $\epsilon$  is essential
(and the effect of  $\epsilon_\omega$ is negligible) then  $f\cong$
constant and the energy spectrum obey Kolmogorov  $5/3$ power law.  If
on the other end of this range  $\epsilon$ is not essential then  $f$
must have the form
\begin{equation}
\label{form}
f \cong (kL_\omega)^{-4/3}
\end{equation}
and consequently
\begin{equation}
E(k) = C\epsilon^{2/3}_\omega k^{-3}
\end{equation}
(where  $C$  is a constant).

For stratified medium Obukov \cite{AMO} introduced the temperature
inhomogeneity dissipation rate
\begin{equation}
\label{rate}
\epsilon_T = 2\chi\int^\infty_0 k^2E_T(k)dk
\end{equation}
where  $E_T$  is the temperature spectra and  $\chi$ is the heat
conductivity of the medium.  He further postulated that the turbulent
component of  $T$  is dependent on this parameter.

For the (stratified) Antarctic medium we would like to enlarge the domain of this
postulate to include the velocity components of the flow.  This enables
us to introduce the buoyancy (length) scale \cite{ASM,OAM}
\begin{equation}
\label{scale}
L_B = (\alpha g)^{-3/2} \epsilon^{5/4}\epsilon_T^{-3/4}
\end{equation}
where  $(\alpha g)$  is the buoyancy parameter.  The existence of this
second length scale for stratified two dimensional flow lead us to
replace (\ref{range}) by
\begin{equation}
\label{flow}
E(k) = f(kL_\omega, kL_B) \epsilon^{2/3}k^{-5/3}
\end{equation}
However since stratification and enstrophy conservation are
independent of each other we infer that  $f$ must have the form
\begin{equation}
\label{each}
f \cong (kL_\omega)^r(kL_B)^s.
\end{equation}
It follows then that the spectral dependence on  $k$  is given by
\begin{equation}
\label{that}
E(k) \sim k^{r+s-5/3}.
\end{equation}
We conclude therefore that various combinations of  $r,s$  are
possible and this will lead to different spectral dependencies on  $k$.

Thus if
$$
E(k) \sim k^{-q}
$$
and the dissipation  $\epsilon$ is negligible we must have then
$$
r + s = 5/3 - q,\;\;\frac{r}{2} + \frac{5}{4} s + \frac{2}{3} =
0
$$
which yields
$$
r = \displaystyle\frac{33-15q}{9},\;\;s = \displaystyle\frac{15
q - 18}{9}.
$$

From the spectral plots for the data under consideration we see that (approximately)
$$
E(k) \sim k^0
$$
for a large segment of the inertial range which is characteristic of
the ``buoyancy range turbulence'' as predicted by Bolgiano \cite{RBJ,BRJ}.

It is interesting to note in this context that Kriachnen
\cite{KR} already observed that the ``energy spectrum of the flow
depends on the details of the nonlinear interaction embodied in the
equations that govern the flow and can not be deduced solely from the
symmetries, invariances and dimensionality of the equations''.

Finally we would like to observe that the data under consideration
contains some discontinuities.  These can change completely the
asymptotic behavior of the spectrum.  To demonstrate this assume that
the data is described by
\begin{equation}
\label{411}
D(x) = CH(x - x_0) + g(x)
\end{equation}
where  $g(x)$  is a smooth function whose Fourier transform (FT)
decays exponentially and  $H(x)$  is the Heaviside function
$$
H(x) = \left\{ \begin{array}{ll}
1   &x \geq 0\\
0  &x < 0.
\end{array} \right.
$$
Differentiating (\ref{411}) we have
\begin{equation}
\label{412}
D^\prime(x) = C\delta(x - x_0) + g^\prime(x)
\end{equation}
and the  FT  of (\ref{412}) is
\begin{equation}
\label{413}
\tilde{D}^\prime(k) = C + \tilde{g}^\prime (k)
\end{equation}
The FT of  $D$  is obtained then by dividing (\ref{413}) by $k$  which
shows clearly that the asymptotic behavior of  $\tilde{D}(k)$  is
proportional to  $k^{-1}$.

We conclude then that a proper filter for the removal of these
discontinuities from the data is needed in order to obtain the true
spectrum of the turbulent residuals.  Such a filtering algorithm is
given by the  $K-L$ decomposition which was described in Sec. 2.

\section{Conclusion}

Using the coherency test advanced by Dewan we are able to characterize
the flow under consideration as one that has the characteristics of
2-d turbulence.  One stumbling block for this interpretation is the
absence of $-3$ slope in part of the inertial range.  To explain this
we introduced a model that takes into account the stratification of
this flow.  This model shows that when buoyancy effects are taken into
account different slopes of  $E(k)$  are possible.  Thus we believe
that we introduced evidence for the interpretation of this spectra as
one belonging to BTR.

\centerline{\bf Acknowledgment}

The author is deeply indebted to Dr. J. Rees and the British Antarctic
Survey Team, Cambridge, UK for access to the
antarctic data and to Dr. J. Rees and O. Cote for bringing to his
attention the peculiar spectrum of this data.

\newpage

\begin{table}[h]
\begin{center}
\begin{tabular}{|l|cccc|crrr|}\hline
   &&   &$m_1$ &&  &$m_2$&& \\ \hline
u at 5m && &2  && &42&&\\
v at 5m && &2&&   &26&&\\
w at 5m && &2 &&  &30&&\\
T at 5m && &4  && &26&&\\ \hline
u at 16m&&  &2&&   &42&&\\
v at 16m &&  &2 &&  &40&&\\
w at 16m&&  &3 &&  &37&&\\
T at 16m && &2 &&  &41&&\\ \hline
u at 32m&&  &4 && &48&&\\
v at 32m && &1&&  &40&&\\
w at 32m&&  &4&&  &51&&\\
T at 32m&&  &2&&  &42&&\\ \hline
\end{tabular}
\end{center}
\end{table}
\centerline{Table 1}

\begin{thebibliography}{www}

\bibitem{VMC} V. M. Canuto, M.S. Dubovikov and D.J. Wielaard - A
dynamical model for turbulence vs Two-dimensional Turbulence,
Phys. Fluids, {\bf 9} p. 2141-2147 (1997).

\bibitem{MEM} M.E. Maltrud and G.K. Vallis - Energy spectra and
choherent structure in forced two-dimenmsional and beta plane
turublence, J. Fluid Mech. {\bf 228} p. 321-342 (1991).

\bibitem{WH} Wendal Horton and A. Hasegawa - Quasi two-dimensional
dynamics of plasmas and fluids, Chaos, {\bf 4}, p. 227-251 (1994).

\bibitem{RK}  R. Kraichnan -  Phys. Fluids, {\bf 10}, p. 1417 (1967).

\bibitem{DKL} D. K. Lily - Numerical Simulation of two-dimensional
turbulence, Phys. Fluid Supp. 2, II-233 (1969).

\bibitem{GKB} G.K. Batchelor - Computation of the energy spectrum in
homogeneous two-dimensional turbulence, Phys. Fluid Supp. 2, II-240 (1969).

\bibitem{JCK} J. C. King, S. D. Mobbs, J.M. Rees, P.S. Anderson and
A.D. Culf.  The stable Antarctic boundary layer experiment at Haley
Station, Weather, {\bf 44}, p. 398-405 (1989).

\bibitem{NRE} N.R. Edwards and S.D. Mobbs - Observation of isolated
wave-turbulence interactions in the stable atmospheric boundary layer,
Q.J.R. Meteorol. Soc., {\bf 123}, p. 561-584 (1997).

\bibitem{EMD} E.M. Dewan - On the nature of atmospheric waves and
turbulence, Radio Sci., {\bf 20}, p. 1301-1307 (1985).

\bibitem{RBJ} R. Bolgiano, Jr. - Turbulent spectra in a stably
stratified atmosphere, J. Geo Res. {\bf 64}, p. 2226-2229 (1959).

\bibitem{BRJ}  R. Bolgiano, Jr. - Structure of turbulence in
stratified media, J. Geo Res. {\bf 67}, p. 3015-3023 (1962).

\bibitem{JJF} F. Einaudi and J.J. Finnigan - Wave turbulence dynamics in the stably
stratified boundary layer, J. Atmos. Sci., {\bf 50}, p. 1841-1864 (1993).

\bibitem{CPMG} C. Penland, M. Ghil and K.M. Weickmann - Adaptive
filtering and maximum entropy spectra with applications to changes in
atmospheric angular momentum, J. Geo. Res. {\bf 96} p. 22659-22671
(1991).

\bibitem{DNA}  D.N. Axford - Spectral analysis of an aircraft
observation of gravity waves, Q.J. Roy. Met. Soc., {\bf 97},
p. 313-321 (1971).

\bibitem{UF}  U. Frisch - Turbulence, Cambridge Univ. Press. (1995).

\bibitem{ASM} A.S. Monin and R.V. Ozmidov - Turbulence in the ocean,
D. Reidal Pub. Co. (1985).

\bibitem{AMO} A.M. Obukhov - Structure of temperature field in
turbulent flow, Izv. Ale. Nauk SSSR, Ser. Geofiz {\bf 13} p. 58-69
(1949).

\bibitem{OAM} A.M. Obukhov - On stratified fluid dynamics, Dokledy AK,
Nauk SSSR, {\bf 145} p. 1239-1242 (1962).

\bibitem {KR} R. Kraichnan - On Kolmogorov inertial-range theories,
J. Fluid Mech. {\bf 62}, p. 305-330 (1974).

\bibitem {WNV} W.N. Venables and B.D. Ripley - Modern applied
statistics with S-plus, Springer-Verlag (1996).
\end{thebibliography}
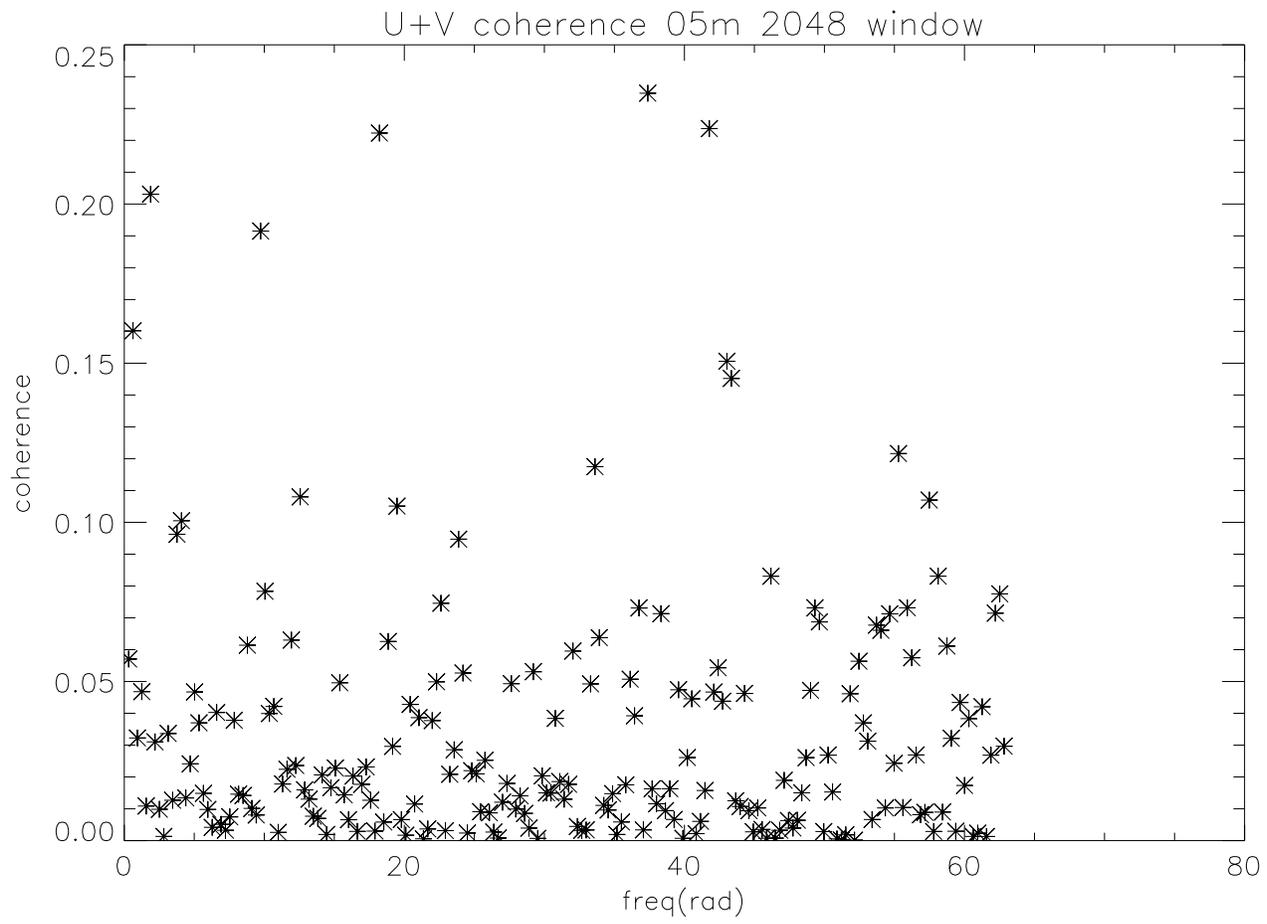
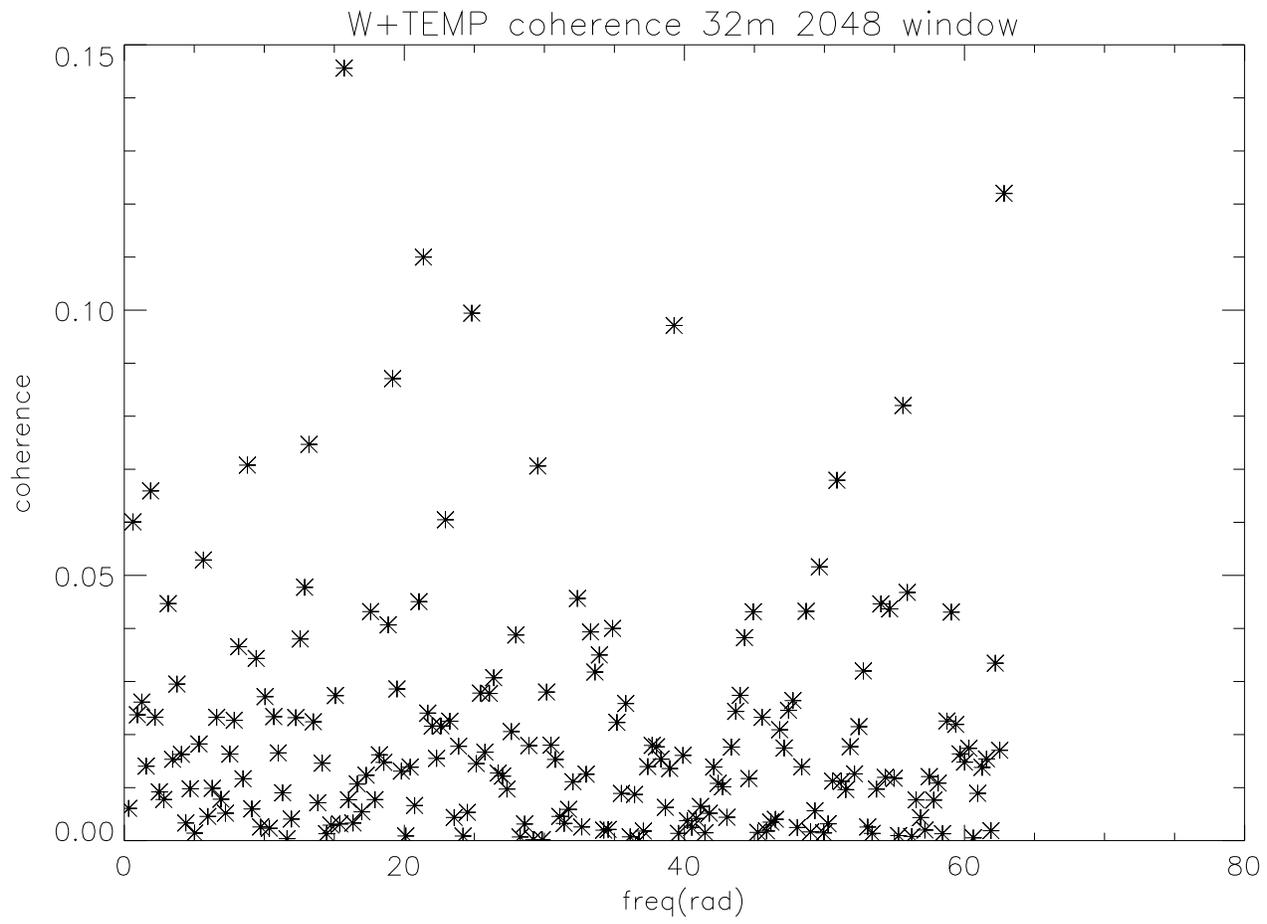
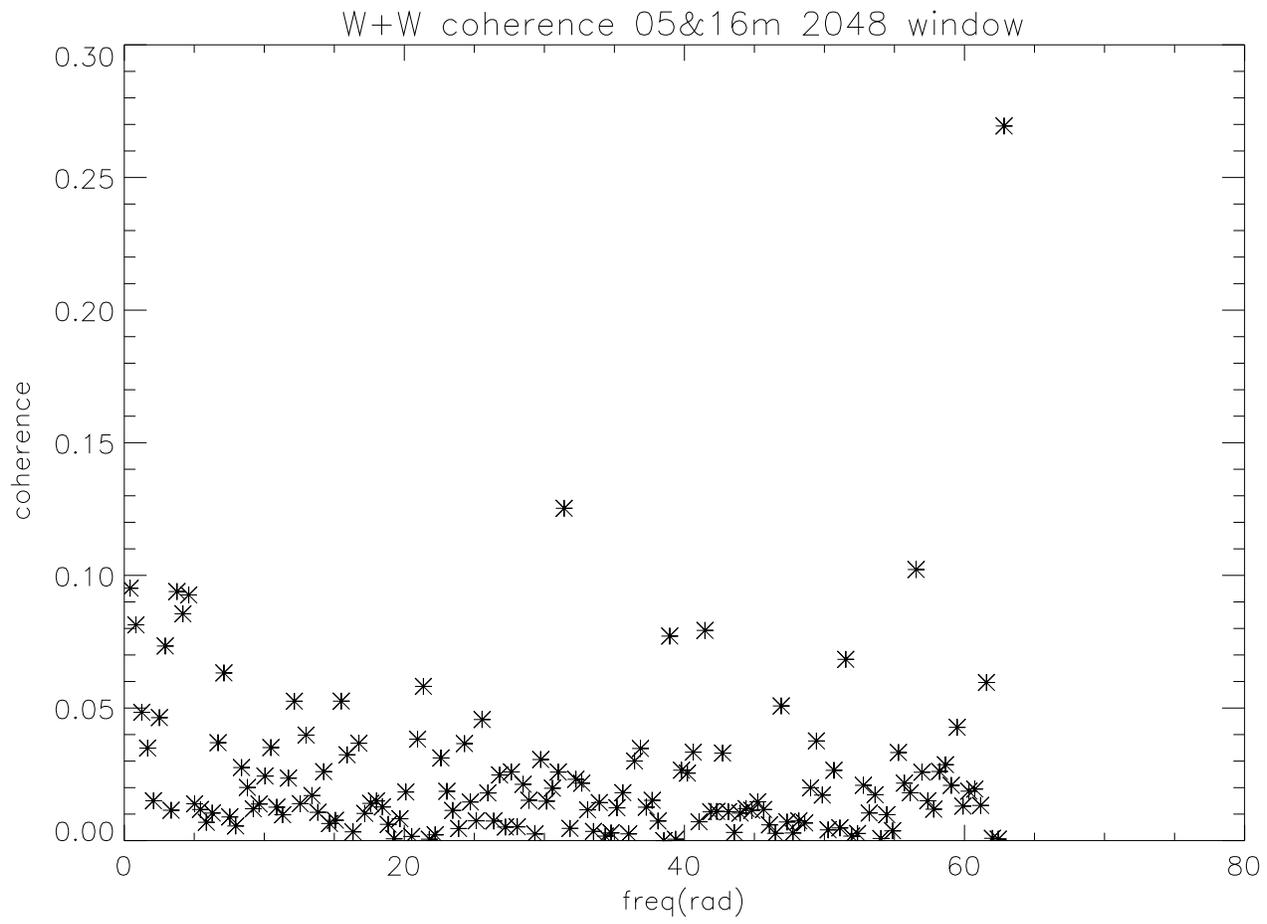
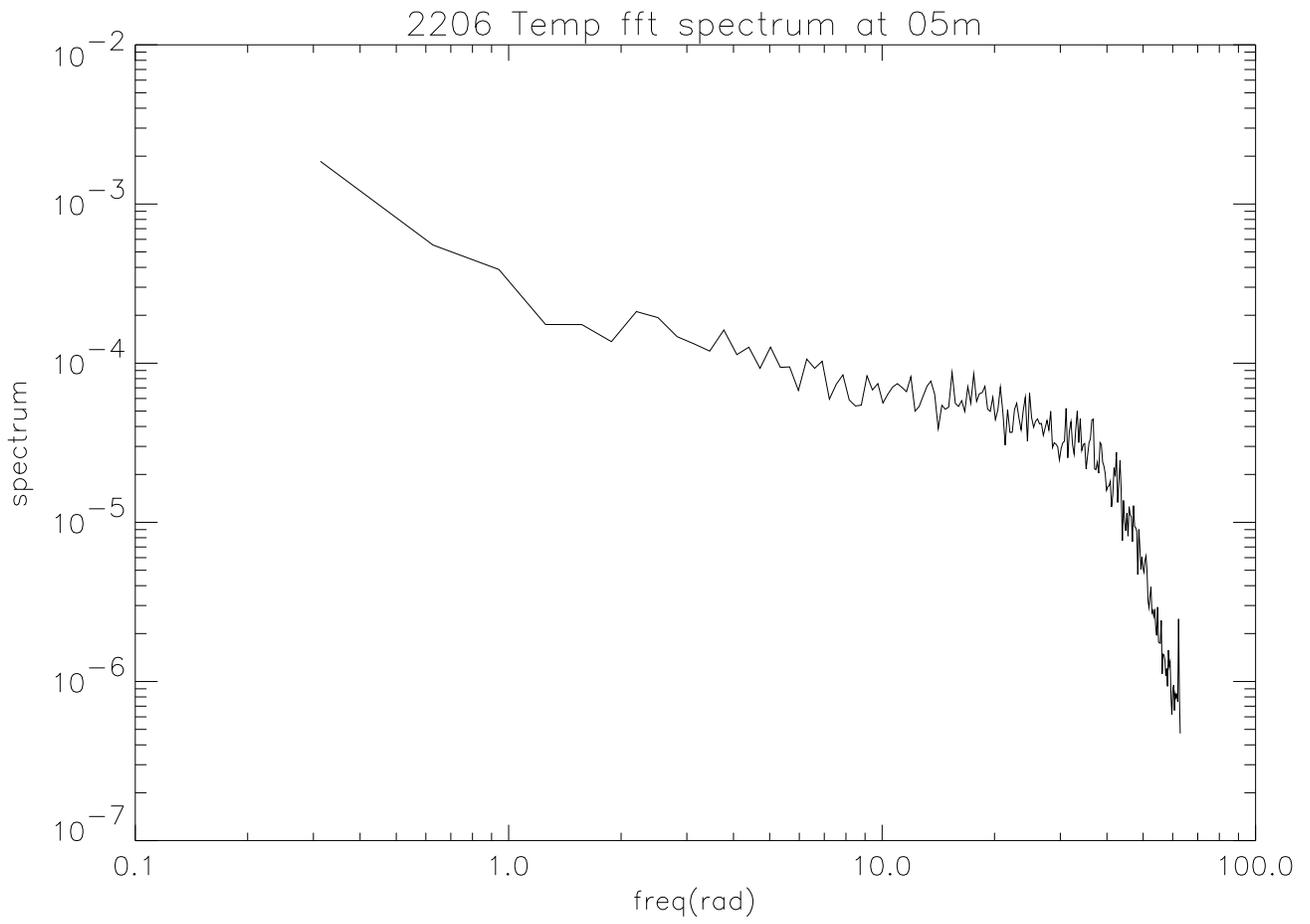
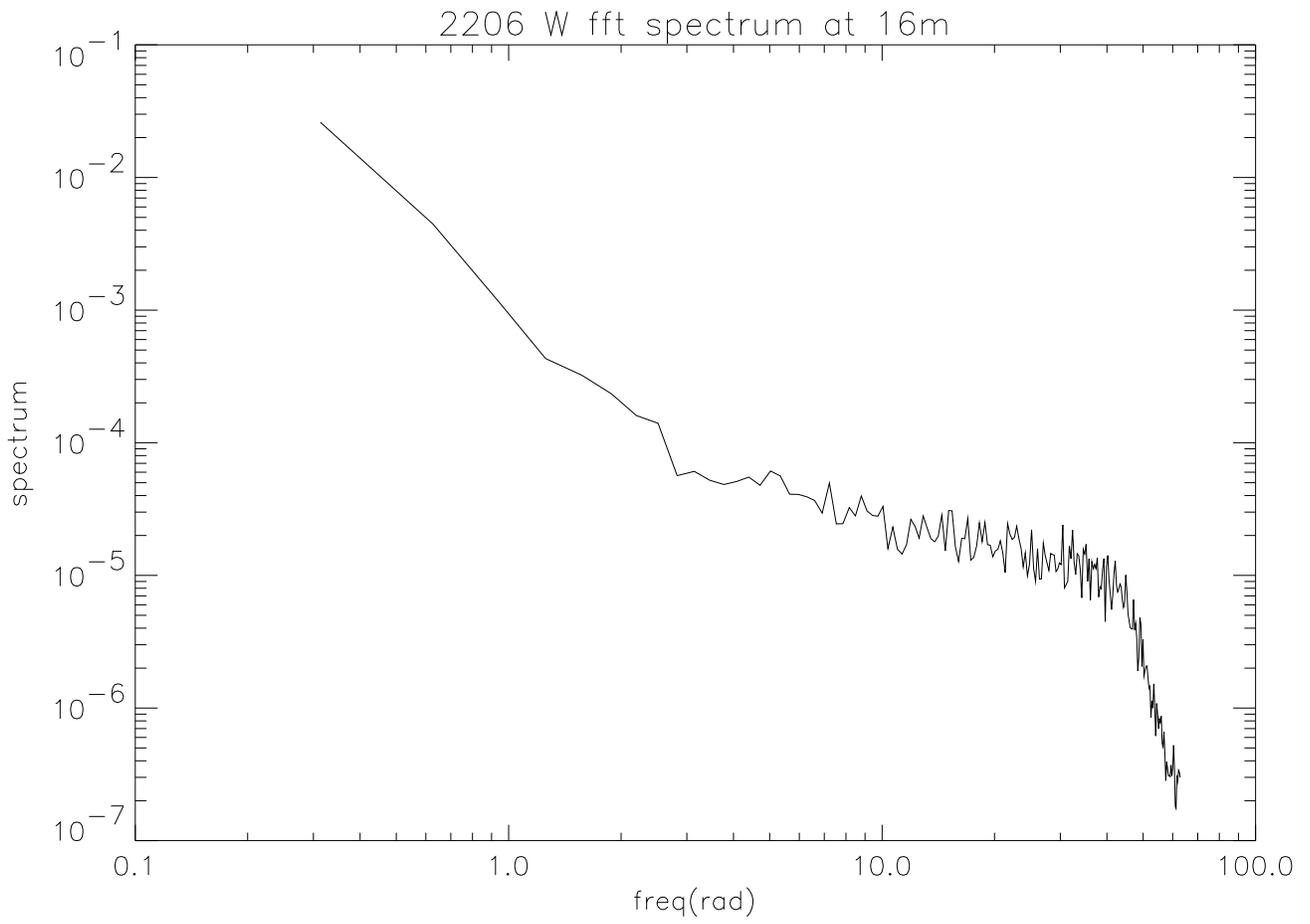
\end{document}